\documentclass[sigconf]{acmart}

\pdfoutput=1

\AtBeginDocument{%
  \providecommand\BibTeX{{%
    \normalfont B\kern-0.5em{\scshape i\kern-0.25em b}\kern-0.8em\TeX}}}

\definecolor{venuec}{HTML}{ef8d22}
\definecolor{defaultc}{HTML}{e5e5e5}

\acmConference[CHI '22]{ACM Conference on Human Factors in Computing Systems}{Apr 30--May 6, 2022}{New Orleans, LA}
\usepackage{color}
\usepackage{float}
\usepackage{array}
\usepackage{textcase}
\usepackage{fancyvrb} % for "\Verb" macro
\usepackage{wrapfig}

\definecolor{class}{RGB}{117, 161, 210}

\usepackage{hyperref}

\setcopyright{none}

\usepackage{multirow}

\usepackage{amsmath}

\usepackage{xspace}
\newcommand{\ie}{{i.e.,}\xspace}
\newcommand{\eg}{{e.g.,}\xspace}
\newcommand{\ea}{{et~al.}\xspace}

\newcommand{\lib}{ComputableViz}

\newenvironment{revised}[0]{%
    \leavevmode\color{black}\ignorespaces
}{}

\acmSubmissionID{}

\begin{document}

\title{ComputableViz: Mathematical Operators as a Formalism for Visualisation Processing and Analysis}
% \title{ComputableViz: Visualization Operation, Processing, and Analysis}
% \title{Computable Visualization: Mathematical Operation, Processing, and Analysis of Data Visualization}
% \title{Towards automated processing and analysis of data visualizations using mathematical operations}
% \renewcommand{\shorttitle}{Short Title}

% \author{Aoyu Wu}
% % \authornote{Both authors contributed equally to this research.}
% \affiliation{%

\author{Aoyu Wu}
\affiliation{%
  \institution{The Hong Kong University of Science and Technology}
  \city{Hong Kong SAR}
  \country{China}
}\email{awuac@connect.ust.hk}

\author{Wai Tong}
\affiliation{%
  \institution{The Hong Kong University of Science and Technology}
  \city{Hong Kong SAR}
  \country{China}
}\email{wtong@connect.ust.hk}

\author{Haotian Li}
\affiliation{%
  \institution{The Hong Kong University of Science and Technology}
  \city{Hong Kong SAR}
  \country{China}
}\email{haotian.li@connect.ust.hk}

\author{Dominik Moritz}
\affiliation{%
  \institution{Carnegie Mellon University}
  \city{Pittsburgh}
  \country{USA}
}
\email{domoritz@cmu.edu}

\author{Yong Wang}
\affiliation{%
  \institution{Singapore Management University}
  \city{Singapore}
  \country{Singapore}
}\email{yongwang@smu.edu.sg}

\author{Huamin Qu}
\affiliation{%
  \institution{The Hong Kong University of Science and Technology}
  \city{Hong Kong SAR}
  \country{China}
}\email{huamin@cse.ust.hk}
\renewcommand{\shortauthors}{Wu, et al.}

\begin{abstract}
% We present the idea of mathematical operations on visualizations for automated processing and analysis.
Data visualizations are created and shared on the web at an unprecedented speed,
raising new needs and questions for processing and analyzing visualizations after they have been generated and digitized.
However,
existing formalisms focus on operating on a single visualization instead of multiple visualizations,
making it challenging to perform analysis tasks such as sorting and clustering visualizations.
Through a systematic analysis of previous work,
we abstract visualization-related tasks into mathematical operators such as union and propose a design space of visualization operations.
We realize the design by developing ComputableViz, a library that supports operations on multiple visualization specifications.
To demonstrate its usefulness and extensibility,
we present multiple usage scenarios concerning processing and analyzing visualization, 
such as generating visualization embeddings and automatically making visualizations accessible.
We conclude by discussing research opportunities and challenges for managing and exploiting the massive visualizations on the web.
\end{abstract}



%%\keywords{datasets, neural networks, gaze detection, text tagging}

%   \includegraphics[width=\textwidth]{picture/overview.pdf}
%   \caption{Overview of \namel{}: (A) We identify four chart components that can be operated; 
%   (B) We present basic chart operations including arithmetic and collection operators;
%   (C) Chart analysis builds upon those operations to extract meaningful information and enable applications,
%   \eg compositing multiple charts and finding charts with fake-data.}
%   \Description{}

\maketitle

\section{Introduction}
% The past years have seen increasing use of data visualizations online such as news, web articles, and social media.
% By representing data graphically,
% visualizations often serve as the main entry point to data for the public.
% With the rapid evolution of information technology,
% visualizations are created and shared on the web at an unprecedented speed.
% According to Google Trends~\cite{GoogleTrends},
% the search interests of ``visualization'' have surpassed that of ``image'' globally since Feb 2020 and keep growing.

By representing data graphically,
visualizations often serve as the main entry point to data for the public.
With the increasing availability and democratization of visualization authoring tools,
a large number of visualizations have been produced and shared on the web.
Correspondingly,
it is argued that data visualizations are becoming a new data format~\cite{wu2021ai4vis}.
There is a growing research interest in exploring techniques used for processing and analyzing visualizations \textit{after they have been generated and digitized}.
For instance,
researchers have started to investigate the problem of style transfer~\cite{harper2017converting, qian2020retrieve} and example-based retrieval~\cite{hoque2019searching} on visualizations.
Besides,
the explosion of visualizations online has led to needs for summarizing and organizing what information has been presented to the public,
especially in high-impact domains such as public health~\cite{zhang2021mapping,lee2021viral}.
Under this trend,
new ideas and problems are emerging, raising the need to organize research work and inform future research.

In this paper,
we formulate this research topic as \textbf{visualization processing and analysis}.
Our formulation is inspired by the well-established \textit{image processing and analysis}~\cite{niblack1985introduction} since visualizations are widely considered as graphic images.
Thus,
visualization processing and analysis concern processing digitized visualizations through an algorithm and extracting meaningful information from visualizations.
In surveying relevant literature,
we observe that a large body of research has moved from operating on a \textit{single} visualization to \textit{multiple} visualizations.
For example,
style transfer involves ``merging'' a content visualization with a style visualization~\cite{gatys2016image},
and example-based retrieval concerns measuring the difference or similarity between visualizations~\cite{hoque2019searching,li2022structure}.

Researchers have contributed many mathematical formalisms for visualizations to generalize and organize research ideas~\cite{lee2019broadening}.
However,
existing formalisms such as algebraic frameworks~\cite{kindlmann2014algebraic} and Draco~\cite{moritz2018formalizing} focus on modelling operations on a single visualization.
There lacks a formalism for operations on multiple visualizations.
This issue results in both conceptual and practical gaps between different research problems and techniques.
For example,
visualization search engines~\cite{hoque2019searching} and visualization sequencing~\cite{kim2017graphscape} seem conceptually separated,
yet they share the core idea of measuring the similarity between two visualization specifications.
This kind of underlying relevance also provides an opportunity for avoiding ``reinventing the wheel'' in the implementation and development of visualization techniques,
% practical software development,
\eg~a reusable function for computing similarities.

To bridge this gap,
we present a unified framework to formalizing visualization processing and analysis by mathematical operators on visualizations.
As an initial step along this direction, we begin by formalizing visualization operations along two basic dimensions,
\ie~\textit{what} components of visualizations (operand) can be operated on and \textit{how} can they be operated on (operator).
We enumerate possible options of operands and operators by referring to existing implementations about visualization primitives (\ie Vega-Lite~\cite{satyanarayan2016vega}) and image operators (\eg scikit-image~\cite{van2014scikit}), respectively.
Based on a systematic analysis of previous work, 
we contextualize the design space to visualizations.
We formally define operands as visualization primitives (\ie data, transform, mark, and encoding) and identify operators, including three binary operators (\ie union, intersection, difference) and advanced operators (\eg sorting and clustering).
This formalism provides a framework that is sufficiently clean and simple for us to describe existing techniques and communicate their relevance and subtleties.

We next describe \lib{}, our proof-of-concept implementation of visualization operations based on Vega-Lite specifications.
Our implementation presumes that visualization images, when applicable, can be transferred to specifications by reverse-engineering techniques~\cite{poco2017reverse}.
Our generic approach converts Vega-Lite specifications into a relational database,
based on which mathematical operators are implemented on the grounds of relational algebra~\cite{codd2002relational}.
Finally, we present several example applications using our implementation.
We show that our solution can replicate and provide more efficient solutions to previous applications.
We also describe new usage scenarios where \lib{} helps explore visualization collections and conduct analysis.
We conclude by discussing the expressive and generative power of our formalism, important extensions to our design and implementation, as well as future work surrounding processing and analyzing visualizations.
In summary, our contributions are as follows:

\begin{itemize}
    \item We propose a formalism for processing multiple visualizations based on mathematical operations and provide a design space of visualization operations through a literature review;
    \item We implement \lib{}, a library for mathematical operations on Vega-Lite visualization specifications;
    \item We present multiple example applications to demonstrate the benefits, usefulness, and extensibility of \lib{}.
\end{itemize}

\section{Related Work}
Our work is related to image processing and analysis,
visualization processing and analysis,
programming libraries for visualizations,
and mathematical framework of visualizations.

\subsection{Image Processing and Analysis}
Digital image processing and analysis have been undergoing vigorous growth in wide-ranging research fields.
While the continuum of image-related research might not have a clear-off boundary,
a useful distinction is often made by considering image processing to be the discipline where both input and output are images~\cite{niblack1985introduction}.
Under this definition,
image processing often builds upon primitive operations such as algebra, mathematical morphology, and signal processing techniques~\cite{haralick1987image, ritter1990image}.
The area of image analysis furthers the continuum by extracting information from images and ``making sense of'' image ensembles.
Inspired by those concepts,
we aim to contextualize operations, processing, and analysis to data visualizations.

Much research has attempted to intertwine image processing and analysis with visualizations.
A large body of work focuses on applying image processing techniques to scientific visualizations such as medical and flow visualizations (\eg~\cite{zhang2018static, hesselink1988digital,ebling2005clifford}).
Different from them,
we concentrate our focus on data visualizations (\ie charts) that are becoming increasingly popular and important on the web.
More importantly,
we delineate visualization operations from image operations,
since direct operations on visualization images are less meaningful than manipulating semantic information such as visual encodings and data~\cite{brosz2013transmogrification,zhang2020dataquilt}.
We introduce a novel framework of visualization operations grounded on database theory and demonstrate its feasibility through proof-of-concept implementation.

% \subsection{Automation for Visualization}

\subsection{Visualization Processing and Analysis}
The idea of visualization processing and analysis has been implicitly practiced in much early research for different goals.
Several systems are proposed to transform a visualization image into another.
For example,
VisCode~\cite{zhang2020viscode} and Chartem~\cite{fu2020chartem} output a visualization image with encrypted information.
Another line of research studies the problem of style transfer,
\ie to blend two visualizations (one containing visual encodings and one containing data) to create new visualization~\cite{qian2020retrieve,harper2017converting}.
% Their implementations are either specific to some visualization types~\cite{qian2020retrieve} or require manual manipulation~\cite{harper2017converting}.
Some other work aims to model the difference between two visualizations as an action for recommending visualization sequences or animations~\cite{kim2020gemini,kim2017graphscape,shi2020calliope, lin2020dziban} or redesigning visualizations~\cite{wu2020mobilevisfixer,chen2021vizlinter}.
% Their approaches apply a difference (an action) to one visualization to generate a new visualization,
% while our method computes the inverse, 
% \ie computing the action between two visualizations.
% \end{revised}

Research on visualization analysis comes into focus with the rapid popularization and accumulation of visualizations online~\cite{wu2021ai4vis}.
Researchers have started to analyze and mine visualization ensembles to derive useful information such as visualization usage online~\cite{battle2018beagle,hoque2019searching} and design patterns~\cite{smart2020color,chen2020composition}.
Another application is the meta-visualization analysis that aims to explore the data encoded in visualization collections (\eg~\cite{xu2018chart,zhao2020chartseer}).

% there lacks a formalism for operations on multiple visualizations.
% This issue results into both conceptual and practical gaps between research works.
% For example,
% we could relate work about visualization search engines~\cite{hoque2019searching} to that about visualization sequencing~\cite{kim2017graphscape} or clustering~\cite{xu2018chart},
% which were previously conceptually separated but share the core problem of measuring the similarity between two visualization specifications.

The above research topics are diverse and seemingly disjoint,
resulting into a conceptual gap that hides the relevance.
We propose mathematical operations as a formalism for operating on multiple visualizations,
thereby bridging research ideas that are previously considered separate.
Through our proof-of-concept implementation,
we present an abstraction of their methods that facilitate code reuse and interoperation.
Through several usage scenarios,
we demonstrate that its scope of application is not limited to the above specialized problems and have huge potentials for future research.

% The above research implements their solutions in a heuristic and case-by-case manner that hardly enables interoperation.
% It is, therefore, desirable to study visualization processing and analysis theoretically and develop a foundation that can serve as a basis for related research.
% We contribute an abstraction of those techniques into a set of mathematical operations such as union and clustering,
% whereby presenting a proof-of-concept implementation of visualization operations.
% Through several usage scenarios,
% we demonstrate that its scope of application is not limited to the above specialized problems and have huge potentials for future research.

\subsection{Programming Libraries for Visualization}
Decades of industry and research efforts have given birth to many programming libraries for creating data visualizations.
One notable theoretical base is Wilkinson’s Grammar of Graphics~\cite{wilkinson2012grammar} that describes a visualization as a combination of layered components such as data, aesthetics, geometric, and facets.
Based on Grammar of Graphics,
researchers have proposed several widely-used programming languages and grammars such as ggplot~\cite{wickham2010layered} and
Vega-Lite~\cite{satyanarayan2016vega}.
Those languages describe data visualizations in a declarative manner,
\ie visualization specifications that contain visualization primitives such as data, visual encodings, and guides.
In this way, specifications are stored and circulated in text-based formats.
Besides,
specifications could be extracted from visualization images through computer vision algorithms (\eg~\cite{yuan2021deep,poco2017reverse}).
Those benefits inspire our implementation,
\ie to develop a proof-of-concept programming language for operating on visualization specifications in Vega-Lite language.

Recent research in the HCI and visualization field has started to develop programming languages for different purposes.
Several languages are proposed to tailor the specifications to specific kinds of visualizations such as tree-based~\cite{li2020gotree}, scientific~\cite{liu2021igscript}, probabilistic~\cite{pu2020probabilistic}, and animated~\cite{ge2020canis} visualizations.
Those methods mainly operate on a single visualization specification that is insufficient for processing and analyzing.
Closely related to our work,
Gemini~\cite{kim2020gemini} recommends a transition path between two visualizations by modeling the pairwise difference,
and \citet{lekschas2020generic} proposed Piling.JS, a library for spatial grouping of multiple visual elements.
We propose a library that leverages  mathematical computations (\eg union and intersection) for operating on multiple visualization specifications and present several example scenarios to demonstrate its capacity and usefulness.

% Different from them,
% we propose a library for operating on multiple visualization specifications with mathematical computation (\eg join and intersection) and present several example scenarios to demonstrate its usefulness.

\subsection{Mathematical Frameworks of Visualizations}
Much research effort has been made on developing mathematical frameworks to facilitate the quantitative theorization of visualization,
which fall into the category of information-theoretic and algebraic frameworks~\cite{chen2017pathways}.
Information-theoretic frameworks~\cite{chen2010information,wang2011information} manifest the usage of information theory to quantify visual information such as entropy,
while algebraic frameworks~\cite{kindlmann2014algebraic} describe the uni-directional mappings from input data, through intermediate data representation, to final visualization.
Those frameworks focus on modelling a single visualization, 
whereas we seek to model mathematical computation over multiple visualizations.
\begin{revised}
Through inverse thinking of algebraic frameworks,
our method
models and reasons about computations on the final visualizations according to the computations over the data and its visual representations.
\end{revised}

The theoretical nature of mathematical frameworks has inspired many methods and applications.
For instance,
information-theoretic frameworks have enabled effective methods for exploring multivariate~\cite{biswas2013information} and volumetric~\cite{ip2012hierarchical} datasets,
while algebraic frameworks influenced research on validity checking of visualization~\cite{mcnutt2020surfacing,qu2017keeping} and design of visualization grammars~\cite{pu2020probabilistic}.
Similarly, we hope that our framework could inspire future research on processing and analyzing visualizations.

% underpinning quantitative theorization in visualization

% In developing theories of visualizations,
% much effort has been made on developing mathematical framework 

\section{Visualization Operations}
\label{sec:operation}
\begin{figure}[!t]
	\centering
	\includegraphics[width=1\linewidth]{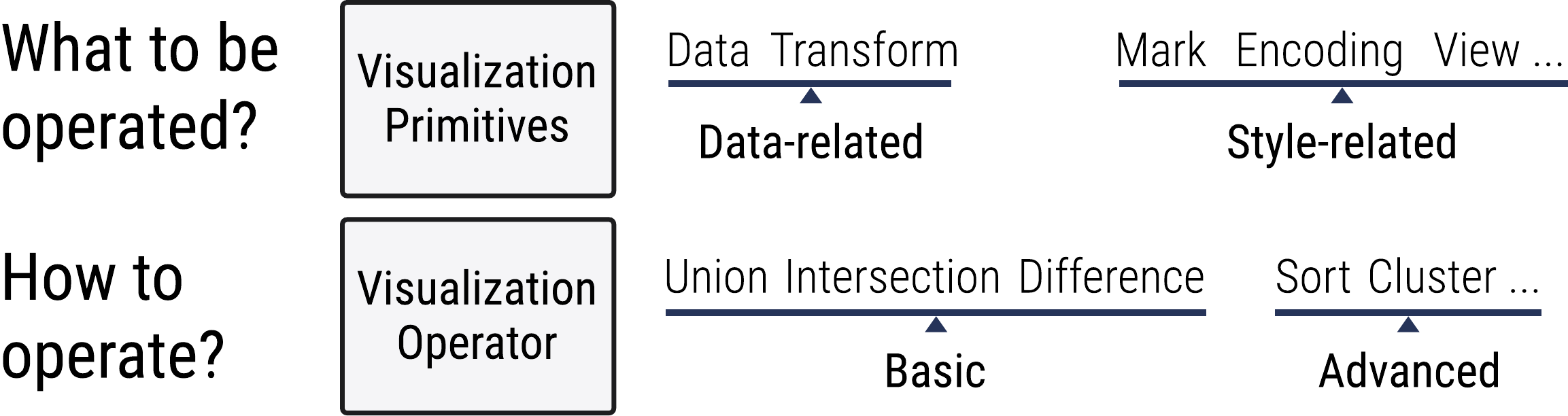}
	\caption{The design space of visualization operations is organised along two dimensions: operation targets (what to be operated) and operation types (how to operate).}
    \Description[]{The design space of visualization operations is organised along two dimensions: operation targets (what to be operated) and operation types (how to operate).}
	\label{fig:chartoperation}
\end{figure}

In this section, we describe visualization operations,
\ie a set of mathematical operations on visualization specifications.
Akin to image operations and processing,
visualization operations are fundamental building blocks to process visualizations.
However,
direct operations on visualization images (\eg arithmetic and morphological operations) might not be always interesting and meaningful.
Instead,
operations on visualizations often involve visualization primitives such as data and visual encodings.
Thus, 
it is desirable to design a set of operations that are tailored to visualizations.

We adopt a deductive approach to conceptualize visualization operations by exploring its design space along two dimensions:
\textit{what} primitives of visualizations can be operated (operation target),
and \textit{how} can they be operated (operation type).
Specifically,
we first list possible candidates along the two dimensions by referring to the existing knowledge in related research.
Then we narrow the range to verify whether each candidate is meaningful,
thereby reaching our conclusions.
Finally,
we valid the design space by systematically reviewing related work concerned with multiple visualizations to understand the coverage.

% In the following text of this section,
% we describe the methodology, results, and validation of the design.

\subsection{Design}
\autoref{fig:chartoperation} illustrates our final design space of visualization operations,
which is organized along two dimensions including operation targets and operation types.
Operation targets are basic visualization primitives that are classified into data-related and style-related. Operation types include basic operations (on two visualizations) and advanced operations (on more than two visualizations).
We do not argue that the design space is comprehensive.
Instead, we consider our design to be initial efforts to conceptualize visualization operations by presenting a basic set.

\subsubsection{Operation Target}
\label{sec:operation:target}
We decide our scope of operation targets by selecting visualization primitives in declarative visualization grammars.
Specially,
we refer to the Vega-Lite~\cite{satyanarayan2016vega} language which has been gaining extensive popularity~\cite{pu2021special}.
At the top level,
a Vega-Lite specification consists of view specification, data/dataset, transform, mark, encoding, view composition, parameters, and configurations.
We classify them into data (including data/dataset and transform) and style (including mark, encoding, and others).

% To manage the complexity of our initial efforts,
% we make a simplifying assumption by focusing on the minimal set of specification primitives,
% including view specification, data/dataset, mark, and encoding.
% Those four types of specification primitives are required,
% \ie missing any of them will result into an invalid visualization.
% \autoref{fig:vegaliteexample} illustrates an example where a Vega-Lite specification is decomposed into those four primitives.

\subsubsection{Operation Type}
% Now that we have considered the operation targets,
% the next problem is to determine possible operation types.
% We obtain a list of candidate operations by consulting relevant documents concerning image operations in digital image processing~\cite{niblack1985introduction} and programming libraries including Google Earth Engine~\cite{googleearth}, Scikit-image~\cite{scikit}, and Matlab~\cite{matlab}.
% Then we filter those operations according to several decisions that are summarized as follows:

Now that we have considered the operation targets,
the next problem is to determine possible operation types.
Following our deductive approach,
we start with obtaining a list of candidate operations by consulting relevant research concerning images and programs,
which are common formats for storing and sharing visualizations~\cite{wu2021ai4vis}.
Since we do not find well-established frameworks for operating on multiple programs,
we consult image-related frameworks concerning image operations in digital image processing~\cite{niblack1985introduction} and programming libraries including Google Earth Engine~\cite{googleearth}, Scikit-image~\cite{scikit}, and Matlab~\cite{matlab}.
Next, we filter those operations according to several decisions that are summarized as follows:

\begin{itemize}
    \item Opting for low-level operations. As a preliminary work, we are interested in identifying low-level operations that can serve as building blocks for high-level operations. Therefore, we limit the scope to low-level operations in image processing characterized by the fact that ``both inputs and outputs are images''~\cite{niblack1985introduction}. That said, we discard operations concerning extracting information such as segmentation and feature extraction.
    \item Excluding spatial- and frequency-based operations. Many image operations aim to improve the image quality by altering the spatial information (\eg morphological operations) or manipulating the frequency domain. However, we find that such direct operations on visualization images are sparse~\cite{brosz2013transmogrification,zhang2020dataquilt} and therefore exclude them.
\end{itemize}

Guided by the above decisions,
we eventually narrow the scope of visualization operations to basic mathematical operations including union, difference, and intersection.
Those visualization operations stem from mathematical operations on images (\eg add and subtract),
while we adapt the wordings in the consideration that a visualization is a collection of visualization primitives. 
We discuss in the following subsection how those basic operations can lead to more advanced operations concerning more than two visualizations.

\begin{figure*}[!t]
	\centering
	\includegraphics[width=1\linewidth]{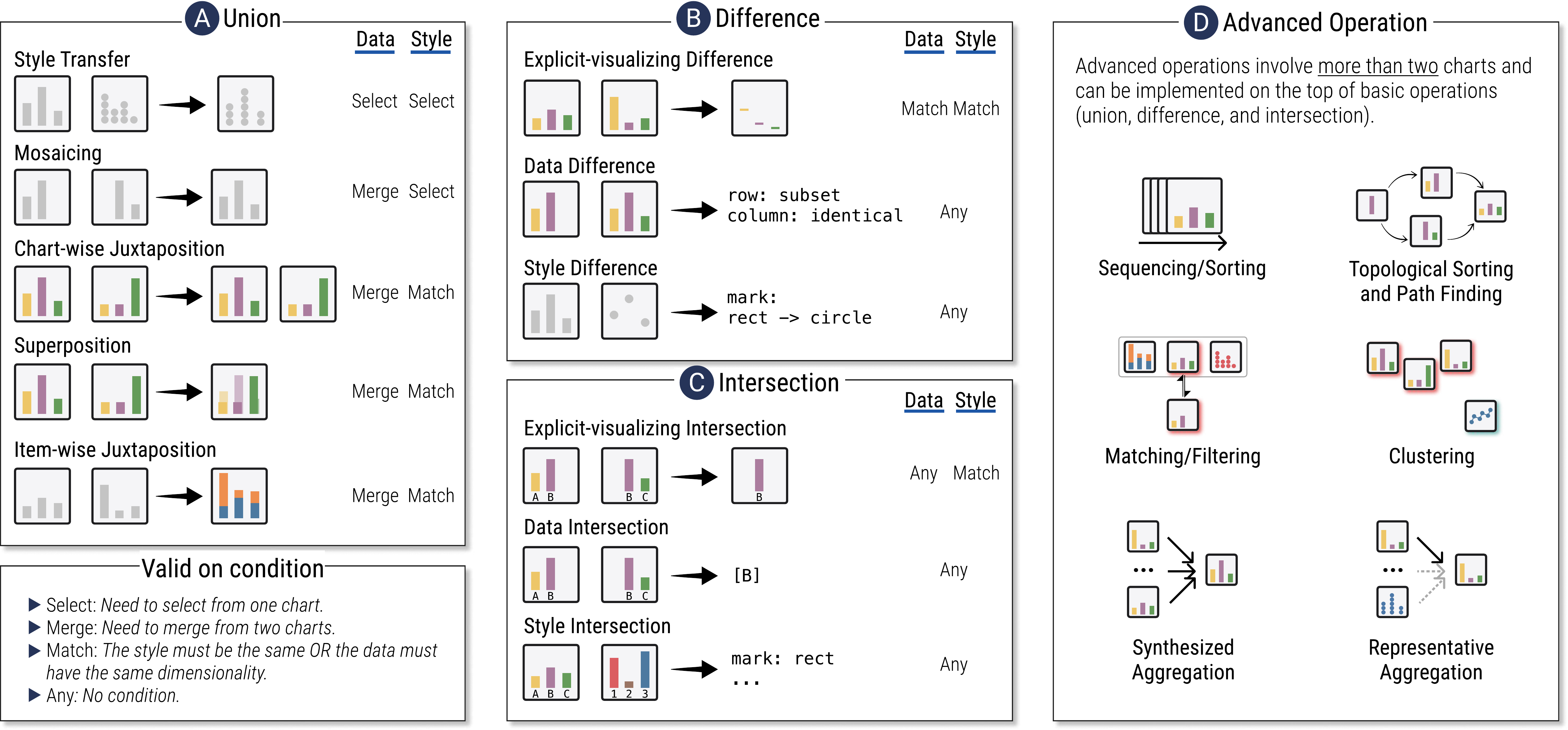}
	\caption{We summarize tasks regarding processing and analysing multiple visualizations from existing literature. We structure tasks according to the basic operations (\ie (A) union, (B) difference, and (C) intersection) and (D) advanced operations. The first three are basic operators on two visualizations, while advanced operations involve more than two visualizations.}
    \Description[]{We summarize tasks regarding processing and analysing multiple visualizations from existing literature. We structure tasks according to the basic operations (\ie (A) union, (B) difference, and (C) intersection) and (D) advanced operations. The first three are basic operators on two visualizations, while advanced operations involve more than two visualizations.}
	\label{fig:operator_overview}
\end{figure*}

\subsection{Refinement}
\label{sec:operation:val}
The design space in~\autoref{fig:chartoperation} is the outcome of the first phrase in our deductive approach,
\ie~starting from a general premise that the fundamental ideas of image processing are legitimate in the context of visualizations.
The next phrase is to refine and contextualise the design space by observation and confirmation. 
Specifically,
we refine and validate visualization operations (\autoref{fig:chartoperation}) by surveying existing work regarding \textit{processing and analysing multiple visualizations}.
Our goal is to construct a design space that is purposeful (\ie solving real problems) and descriptive (\ie generalizing across different studied problems).

% We validate the design space of visualization operations (\autoref{fig:chartoperation}) by surveying existing work regarding \textit{processing and analysing multiple visualizations}.
% Our goal is to investigate whether our design is purposeful (\ie solving real problems) and descriptive (\ie generalizing across different studied problems).

\textit{Collecting Literature.}
We collect the literature by referring to recent related surveys~\cite{wu2021ai4vis,wang2020applying} on the use of AI and machine learning in data visualizations.
We further augment the corpus with related concepts concerning multiple visualizations such as visual piling~\cite{lekschas2020generic}, visual comparison~\cite{lyi2020comparative}, composite visualization~\cite{javed2012exploring,yang2014understand},
and visualization combinations~\cite{schulz2015preset}.
We focus on research about lower-level operations,
excluding statistical analysis of visualization collections (\eg~\cite{battle2018beagle}).

\textit{Extracting Tasks.}
We extract tasks from the surveyed literature.
We mainly adopt the naming of tasks in literature and in few cases create new names when necessary.
For example,
the task ``topological sorting and path finding'' is abstracted from several visualization recommender systems~\cite{kim2017graphscape,lin2020dziban,vartak2015seedb,mafrur2018dive,wu2020mobilevisfixer,qian2020retrieve} that contains a topological graph structure,
where each node represents a visualization and each edge denotes an edit operation.
Those systems aim to find a series of visualizations (nodes) or operations (links) that abstractly are a path. 

\textit{Organizing and Abstracting Tasks.}
We organize those tasks by abstracting them into operations,
\ie which operation can support accomplishing the task.
As illustrated in~\autoref{fig:operator_overview},
we observe two types of scenarios.
First,
\autoref{fig:operator_overview} (A-C) represents tasks that can be supported by basic operators, \ie union, difference, and intersection, respectively.
Each task concerns the data and/or style of the visualizations.
We notice and describe conditions on which the task is valid.
Second,
\autoref{fig:operator_overview} (D) shows several advanced operations concerning more than two visualizations,
which can be developed based on the three basic operators.
We discuss details in section~\ref{sec:operation:valid:advanced}.

\subsubsection{Basic Operation}
\label{sec:ope:basic}
We identify three basic operators including union, difference, and intersection.
\paragraph{Union}
The union operation combines visualization primitives from two or more visualizations.
Union provides support for \textit{style transfer}~\cite{qian2020retrieve, wang2019visualization, harper2017converting} by selecting data from one visualization and selecting styles from another visualization.
A similar task is extending static visualizations with new incoming data~\cite{chen2020augmenting, chen2019towards},
which requires merging the data from two visualizations.
We refer to the general task as \textit{mosaicing},
\ie two or more visualizations of different views of a dataset can be mosaiced to represent the complete view.
We borrow this concept from similar ones such as image mosaicing~\cite{capel2004image} and document mosaicing~\cite{zappala1999document}.

Another set of tasks supported by union is for visualization composition~\cite{lyi2020comparative} and visual comparison~\cite{javed2012exploring,yang2014understand},
including \textit{visualization-wise juxtaposition}, \textit{superposition}, and \textit{item-wise juxtaposition}.
Those tasks merge the data between the visualizations and require that the visual encodings of two input visualizations are equivalent.
It should be noted that visualizations with different visual encodings can still be superimposed or juxtaposed,
which might not always be meaningful and therefore excluded in our discussion.

\paragraph{Difference}
The difference operation can be used to \textit{explicitly visualize difference},
a common approach for visual comparison~\cite{lyi2020comparative}.
This task requires that the data and style of two visualizations should match each other.
If not matching,
one could compute the \textit{data difference} and \textit{style difference}.
Both types of differences can be mapped to semantic operations which are useful in visualization authoring, 
\eg data differences can be used to generate data stories~\cite{shi2020calliope},
and style differences help in recommending~\cite{kim2020gemini,kim2017graphscape,lin2020dziban} or redesigning visualizations~\cite{wu2020mobilevisfixer,chen2021vizlinter}.
However,
there exists many approaches to describe the difference such as natural language~\cite{yu2019flowsense}, parameterized function~\cite{wu2019visact}, and numerical scores~\cite{kim2017graphscape}.
Our goal is to represent the difference in a generic format that can be converted into other formats with advanced techniques.

\paragraph{Intersection}
The intersection operation is highly related to the difference operation.
Therefore,
the task covered by intersection is an analogy of that of difference,
including \textit{explicit-visualizing intersection}, \textit{data intersection}, and \textit{style intersection}.

\subsubsection{Advanced Operation}
\label{sec:operation:valid:advanced}
We identify six tasks concerning advanced operations in surveyed literature. 
Those operations are named ``advanced'' since they concern with more than two visualizations and can be implemented by basic operations.

The first task is \textit{sequencing/sorting},
which is used to find an optimal sequence of visualizations~\cite{shi2019task} or sort visualizations in browsers~\cite{chen2020composition}.
However,
the relationships among visualizations are often non-linear,
but instead modelled using a topological graph structure,
where each node represents a visualization and each edge denotes an edit operation or the visualization difference.
This issue motivates the second task \textit{topological sorting and path finding} that aims to find an optimal destination node given an input node.
This task is widely studied in visualization recommender systems~\cite{kim2017graphscape,lin2020dziban,vartak2015seedb,mafrur2018dive,wu2020mobilevisfixer,qian2020retrieve}.
Both tasks require comparing two visualizations that can be implemented via difference or intersection.

Another two tasks, \ie~\textit{matching/filtering} and \textit{clustering}, are related,
both concerning measuring the difference between visualizations.
Matching/filtering is concerned with visualization retrieval system~\cite{hoque2019searching, saleh2015learning},
\ie to find a visualization that is similar or related to the input visualization.
Clustering provides support for meta-visualization analysis~\cite{zhao2020chartseer, xu2018chart}.
Both tasks might require a distance function that converts the difference/intersection into a numeric score.

Lastly,
\textit{synthesized aggregation} and \textit{representative aggregation} are common methods to summarize multiple visualizations~\cite{lekschas2020generic}.
According to the taxonomy by Lekschas \ea~\cite{lekschas2020generic},
another type of aggregation method is abstract,
\ie to provide a simplistic or schematic representation.
We exclude this type because its output is no longer a visualization but instead the ``category or type'' that needs high-level operations such as visualization classification.
Synthesized aggregation can build upon mosaicing,
while representation aggregation can be implemented via sorting.

\subsubsection{Discussion of Limitations}
From our analysis,
we identify several limitations of visualization operations including the operand (\ie what to operate) and the operator (\ie how to operate).

\begin{figure*}[!t]
	\centering
	\includegraphics[width=1\linewidth]{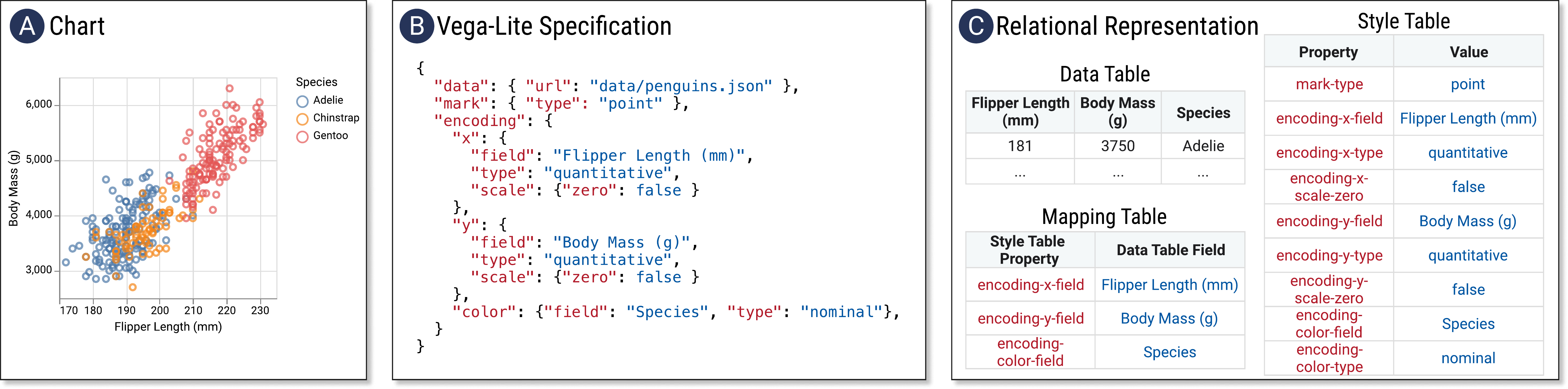}
	\caption{Illustration of visualization specifications and our representation: (A) An example of a scatterplot; (B) Its Vega-Lite specification; and (C) our relational representation. We represent visualization specifications as a relational database including a data table, a style table for storing specifications, and a mapping table describing the relationships between the data table and the style table.}
	\label{fig:vegaliteexample}
    \Description[]{We represent visualization specifications as a relational database, whereby enabling operations grounded on relational algebra such as joins.}
\end{figure*}

\paragraph{Coverage of operands.}
We observe that operands represent a fundamental difference between image and visualization operations.
Image operations directly manipulate the image data (\eg pixel values in RGB),
while operations on visualizations are mostly performed on the specifications (\eg data and encoding).
\begin{revised}
However,
we find few approaches that directly alter the visualization images such as transmogrification~\cite{brosz2013transmogrification}.
As such,
our method does not support operations on the images,
which require further extension.
\end{revised}

\paragraph{Coverage of operators.}
Most of image operators stay legitimate in the context of visualizations, with few exceptions.
We discuss the coverage of our visualization operators from two perspectives: breadth and depth.

From the perspective of breadth,
there exist tasks that are not covered by the operators in~\autoref{fig:operator_overview}.
For instance,
some types of hybrid visualization such as nesting~\cite{javed2012exploring} are performed by replacing one visualization by nested instances of another,
which might be abstracted into a convolution operator.
The F3 system~\cite{355bf403d44a48e6a6cf39ed55cb313e} 
supports creating a view matrix that combines the views’ dimensions by specifying diagonal views,
which can be described by a Cartesian product operator.
Besides,
the current space of mathematical operators cannot precisely describe complex tasks such as interpolating~\cite{schulz2015preset} or morphing~\cite{ruchikachorn2015learning} existing visualizations,
which might require geometric operators.
This problem warrants future research to investigate the design space of hybrid, nested, and coordinated visualizations.
Finally,
our operators exclude unary ones, 
which can help describe actions, interactions, and other provenance data on a single visualization (\eg~VisAct~\cite{wu2019visact} and Trrack~\cite{cutler2020trrack}).
Thus,
an extension to unary operators requires surveying the broad literature such as visualization authoring, interaction, and provenance.

From the perspective of depth,
mathematical operations can be insufficient in fully solving challenging tasks such as visualization clustering,
which needs more advanced approaches such as machine learning (\eg~\cite{zhao2020chartseer}).
However,
mathematical operations provide sensible baseline approaches and facilitate data processing that prepares machine learning approaches.
We demonstrate in~\autoref{sec:chartPA} how mathematical operations can be extended to machine learning tasks.

\section{Implementation}
We develop a proof-of-concept implementation of visualization operations to realize the design in~\autoref{fig:chartoperation}.
Specifically,
we present a python library for operating on Vega-Lite specifications.
Our implementation is mainly grounded on database theory.
We first convert a specification into a relational database and then leverage relational algebra~\cite{codd2002relational} (\eg join) to perform operations.
In the following text of this section,
we first describe our design considerations, followed by implementation details.

% Our proof-of-concept implementation makes one simplifying assumption.
% First, we operate on four visualization primitives as described in section~\autoref{fig:chartoperation}.
% Second, we only handle visualizations which encode a single tabular dataset,
% excluding visualizations such as node-link graphs.

\subsection{Implementation Consideration}
Based on the above analysis of literature,
we derive the following considerations that guide our implementations.

\textbf{C1: Operating visualization primitives separately}.
Our analysis suggests multiple scenarios with different conditions on data and other visualization primitives.
For example,
users might be only interested in computing the difference of style-related primitives between visualizations.
Therefore,
those primitives should be operated separately.

\textbf{C2: Supporting different tasks by passing parameters}.
One operation might support different tasks,
\eg the union operation can be used to achieve style transfer and mosaicing which concern selecting and merging data, respectively.
Therefore,
it is desirable to implement abstract functions with parameterization that allows for code reusability.

\textbf{C3: Representing visualization difference and intersection in interoperable formats}.
Visualization difference and intersection are often represented in different formats (\eg natural language and numeric score) depending on the specific needs.
Our goal is to represent them in interoperable formats which can be transferred to other formats with advanced techniques such as machine learning.

\subsection{Implementation Detail}
We implement visualization operators based on Vega-Lite specifications.
However, 
Vega-Lite specifications are in JSON schema that is semi-structured and difficult to perform operations such as union~\cite{liu2014json}.
We overcome this challenge by converting specifications into a relational database,
whereby implementing operators on the grounds of relational operations.

% \subsubsection{Preprocessing}
\subsubsection{Converting Specifications into Relational Representations}
We convert the visualization specifications into a relational database composed of basic tables and a mapping table. 

\begin{figure}[!t]
	\centering
	\includegraphics[width=1\linewidth]{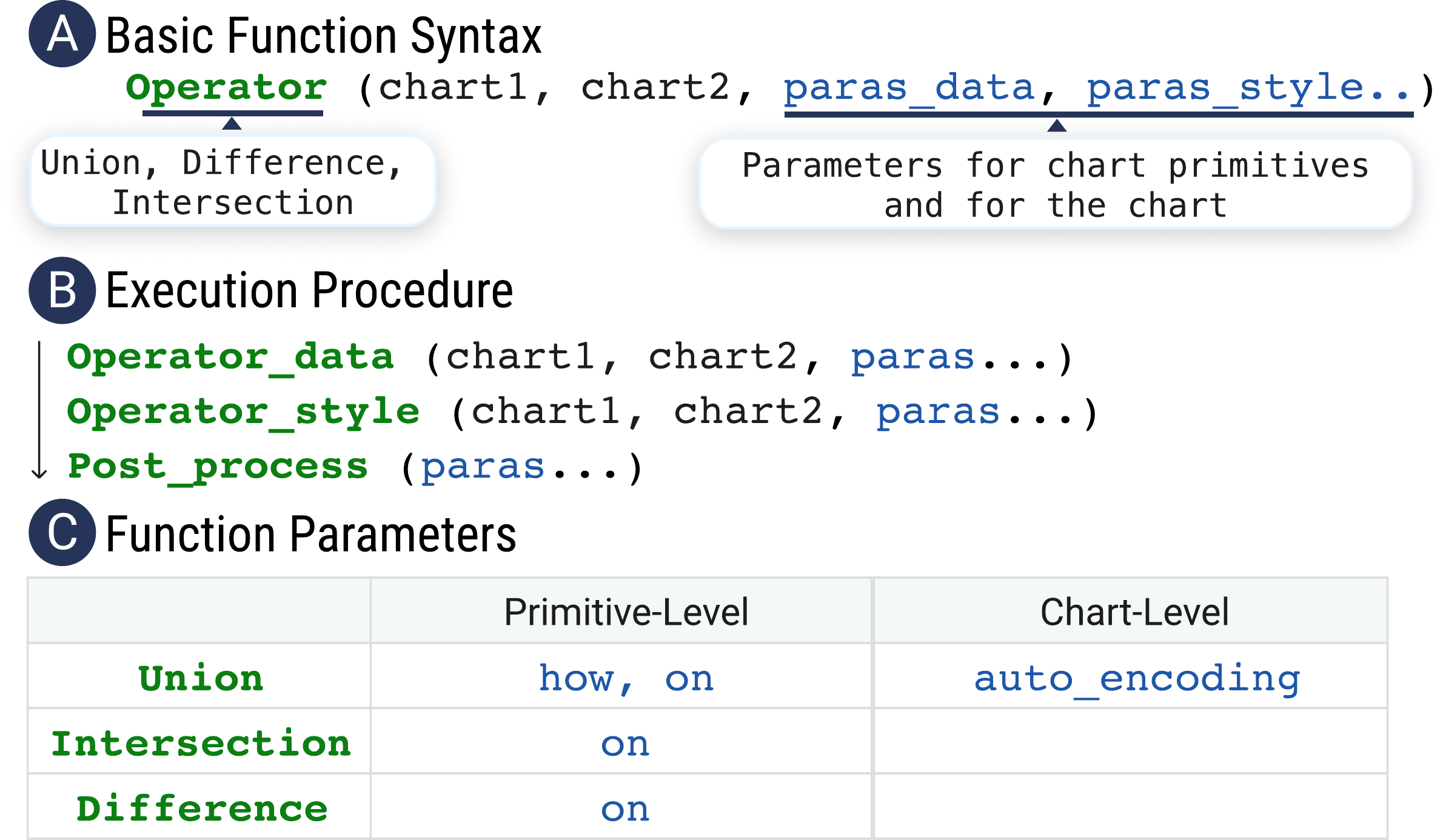}
	\caption{The function argument syntax of visualization operators: (A) The parameters of visualization operators include two input visualizations, primitive-level and visualization-level parameters; (B) The operator works on visualization primitives separately, followed by a post-processing step; (C) The primitive-level and visualization-level parameters vary according to the operation.}
    \Description[]{(A) The parameters of visualization operators include two input visualizations, primitive-level and visualization-level parameters; (B) The operator works on visualization primitives separately, followed by a post-processing step; (C) The primitive-level and visualization-level parameters vary according to the operation.}
	\label{fig:implementation}
\end{figure}

\begin{figure}[!t]
	\centering
	\includegraphics[width=1\linewidth]{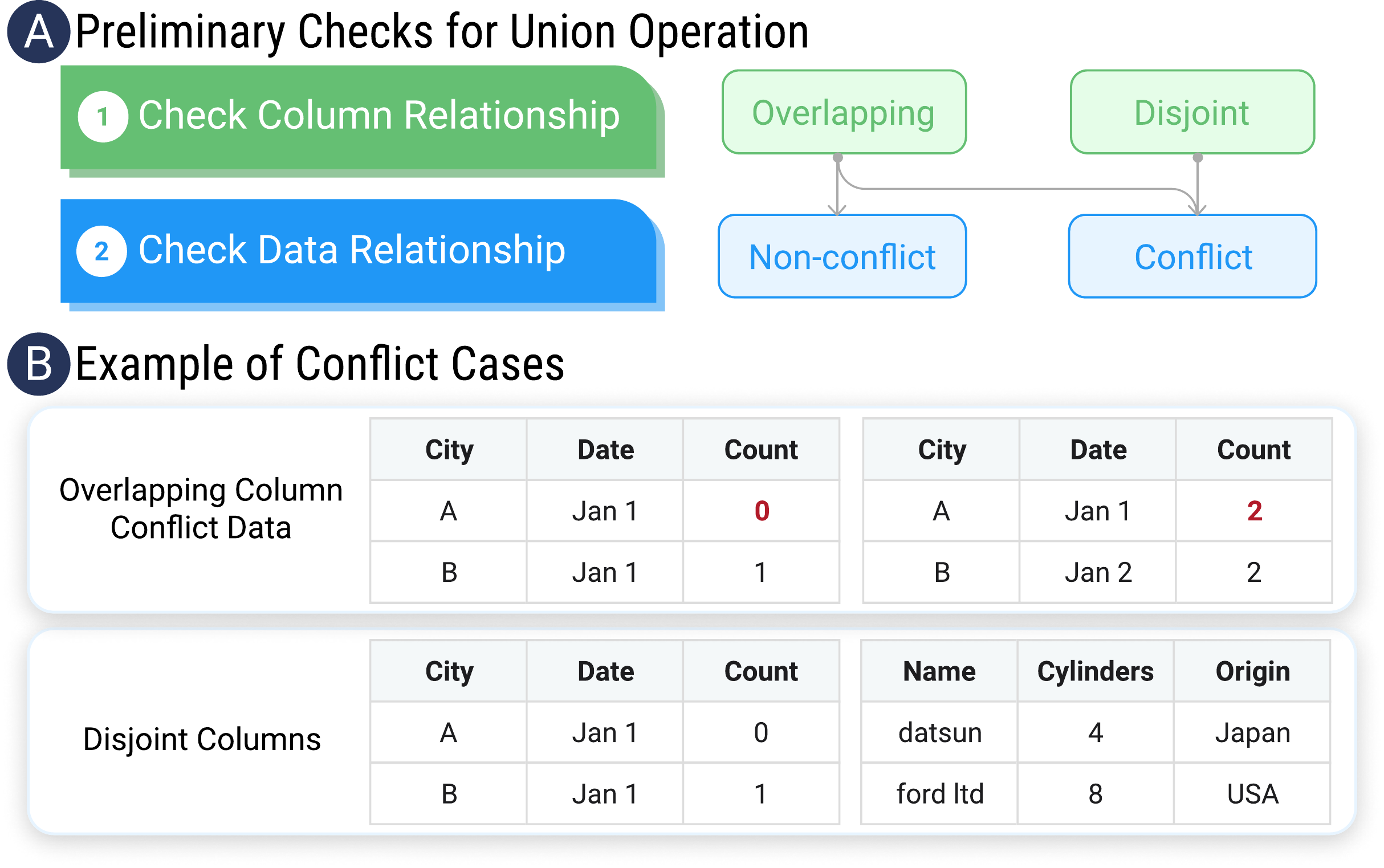}
	\caption{Mechanism for handling data conflicts in the union operation: (A) The union operation starts with preliminary checks; (B) Conflicting data tables cannot be automatically combined.}
    \Description[]{(A) The union operation starts with preliminary checks; (B) Conflicting data tables cannot be automatically combined.}
	\label{fig:union}
\end{figure}

\begin{figure*}[!t]
	\centering
	\includegraphics[width=1\linewidth]{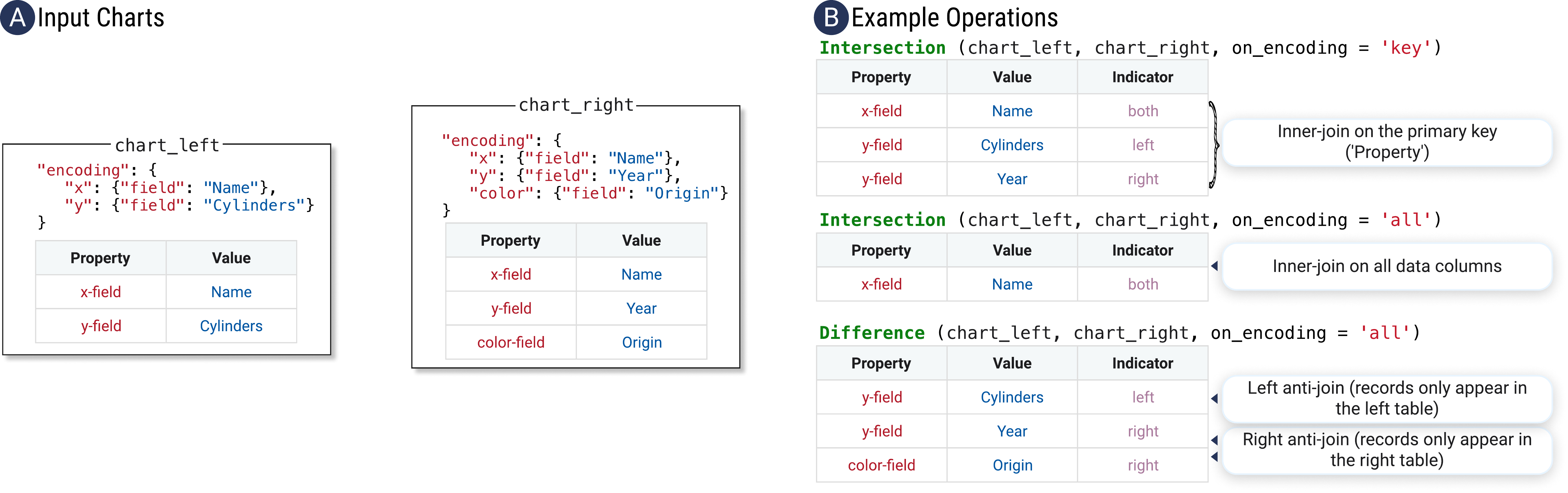}
	\caption{Example of operations: (A) The encoding table of input visualizations; (B) Examples of intersection and difference operations.}
    \Description[]{(A) The encoding table of input visualizations; (B) Examples of intersection and difference operations.}
	\label{fig:example_operation}
\end{figure*}

\textit{Basic Tables.}
As shown in~\autoref{fig:vegaliteexample} (C),
our relational representation consists of tables for each visualization primitive.
The data table is mostly consistent with the original data.
However,
our relational representation drops data columns (fields) that are not encoded to avoid redundant information by default.
For style-related primitives,
we convert the JSON format to property-value table structure.
We flatten nested JSON objects by concatenating keys with nested levels using a dash character \begin{revised}(\eg ``encoding-x-field'' in~\autoref{fig:vegaliteexample})\end{revised}
which is reversible since the JSON scheme does not allow duplicate keys.

\textit{Mapping Table.}
We implement a mapping table to store the relationships between the data table and other tables such as encoding.
Those mapping relationships need to be maintained when changing the data.

\textit{Alternative Representations.}
We consider multiple alternative representations of specifications including logic programming (\eg Draco~\cite{moritz2018formalizing}),  relational programming (\eg~VQL~\cite{luo2020interactive} and CompassQL~\cite{wongsuphasawat2016towards}), and set-theoretic programming~ (\eg Viser~\cite{wang2019visualization}).
However,
those specifications are lossy and contain limited information.
For instance,
Draco expresses data by a set of data attributes instead of the entire dataset,
and VQL discards non-data encodings such as view sizes and interactions.
Different from them,
our relational representation is lossless and reversible to the original Vega-lite specifications.
% Different from them,
% our relational representation is reversible to the original Vega-lite specification and therefore has the same expressive power with Vega-Lite.

\subsubsection{Implementing Visualization Operation by Relational Operators}
\autoref{fig:implementation} illustrates the functions implemented for visualization operations.
The function syntax takes two visualizations and parameters as input.
Each operation is first executed on the data table and the style table separately, followed by post-processing (\textbf{C1}).
We implement the operation on those tables based on the JOIN clause in relational algebra~\cite{codd2002relational}.
Specially,
the union operator builds on the \verb+FULL OUTER JOIN+ clause which merges all records when there is a match in the left table or the right table records.
Similarly,
we realize the intersection operator by the \verb+INNER JOIN+ clause that requires records to match one in both tables.
In a slightly different manner,
the difference operator consists of three steps,
including the left and right \verb+ANTI-JOIN+ that find records that only appear in either the left or the right table.

\paragraph{Function Parameters}
As shown in \autoref{fig:implementation} (C),
the function parameters vary depending on the operator (\ie union, intersection, or difference) and the level (\ie visualization-level or primitive-level),
that are listed as follows (\textbf{C2}).

The \verb+on+ parameter is a common one for primitive-level operations.
It specifies columns on which the \verb+JOIN+ clause is performed.
Specially,
it can take either \verb+'key'+ (\ie joining on primary keys) or \verb+'all'+ (\ie joining on all columns).

The \verb+how+ parameter handles the data-conflicting scenarios as in~\autoref{fig:union} (B).
Its value could be \verb+'left'+, \verb+'right'+ (\ie selecting either the left or right one) or \verb+'merge'+ (\ie merging two tables by adding a new indicator column denoting whether the record is from the left or right table.

The last parameter \verb+'auto_encoding'+ switches a post-processing step for the union operation.
If true,
it will automatically create a new visual encoding for indicator columns.
An indicator column consists of two constant values for data records in the left and right visualization, respectively.
Here we implement a heuristic to assign new visual encoding channels,
\ie to pick an idle channel (if any) in the order of color, opacity, column, and row.

\paragraph{Union}
We perform preliminary checks on the input data tables following two steps, as shown in~\autoref{fig:union}.
First,
we check whether the columns of two data tables are not disjoint,
since \verb+FULL OUTER JOIN+ requires common columns on which tables are merged.
Second,
we detect conflicting data records between two tables using a heuristic-based method.
We compare data records according to their primary keys (\ie columns that contain values that uniquely identify each record).
However,
it still remains an unsolved problem to automatically detect primary keys~\cite{jiang2019holistic}.
Therefore, 
we make a simplifying assumption that a primary key is a unique key of ordinal, categorical, or temporal data types.
\autoref{fig:union} (B) illustrates an example where data conflicts are detected according to the primary key~\verb+<City, Date>+.

The last post-processing step handles the problem of broken links after the union operation.
Specially,
some values in the encoding and transformation specification (\eg \verb+field, groupby+) are linked to corresponding data columns,
which might be missing in the merged data tables.
Therefore,
the last step recovers the links by assigning new data columns to those missing links.
Our assigning strategy prioritizes data columns with the same data type of the origin one.

\paragraph{Intersection and Difference}
We represent the results of intersection and difference in database formats that are common and convenient to process (\textbf{C3}),
as shown in~\autoref{fig:example_operation}.
We implement both operators using the \verb+JOIN+ relational operator.
This operator creates a new column named \verb+indicator+ that has a categorical type with the value of ``left'' if the record only appears in the left visualization, ``right'' if the record appears in the right visualization, or ``both''.

Our representations provide basic information that can be converted to other practical formats.
For instance,
the difference operation (\autoref{fig:example_operation} (B-bottom)) can be used to detect value conflicts by finding duplicates on the property column.
It is also possible to transform the results into natural language sentences using template-based methods (\eg ``adding a new encoding by assigning the data field Year to the encoding channel x'').
Moreover,
distance functions can be proposed to map the table into a numeric score that enables advanced operations such as clustering and matching.

\section{Usage Scenarios}
\label{sec:chartPA}
To demonstrate the usability and extensibility of visualization operations,
we present several usage scenarios regarding visualization processing or analysis.
We situate those scenarios in the context of existing literature and illustrate how our implementation offers new and more efficient solutions to those problems.
The source code is provided as supplemental material.

We present a total of six usage scenarios. 
The first two scenarios focus on visualization processing including creating accessible visualizations and compositing visualizations.
Both scenarios are implemented by a single-line union operation,
showing that our abstraction can cohere previously separate ideas and promote code reuse.
The last four scenarios further the application to visualization analysis.
We show that our operations can be combined or extended to support version control, perform meta-visualization analysis, conduct clustering analysis, and explore the ``genealogy'' tree of visualizations.
Those scenarios demonstrate that our formalism and implementation can generate new, feasible solutions to emerging problems in a more automated and scalable manner.

% Our analysis in~\autoref{sec:operation} shows that our formalism of visualization operations can abstract and describe existing approaches for visualization processing and analysis from a conceptual perspective.
% In this section,
% we present several usage scenarios to demonstrate the usability and extensibility of~\lib{} from a practical perspective.
% We situate those scenarios in the context of existing literature and illustrate how our implementation offers new and more efficient solutions to those problems.
% The source code is provided as supplemental material.

\begin{figure*}[!t]
	\centering
	\includegraphics[width=1\linewidth]{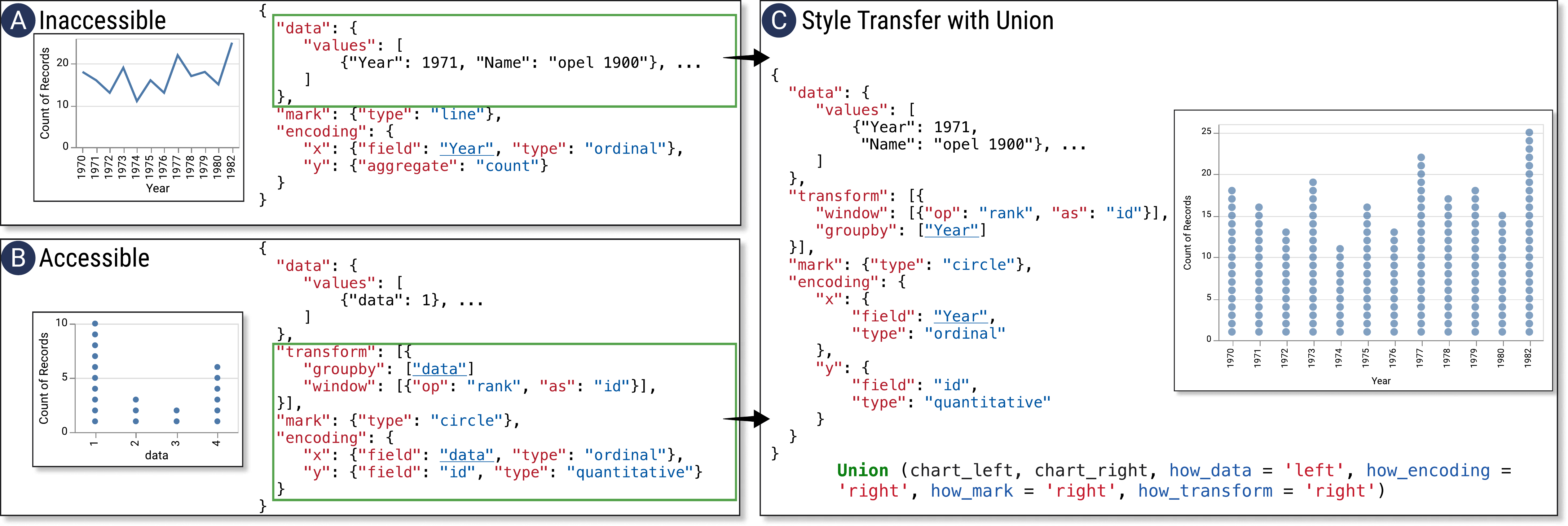}
	\caption{ The union operators enables creating accessible visualization automatically:
	(A) A original line visualization that is inaccessible for people with intellectual and developmental disabilities;
	(B) A dot-based plot that is accessible;
	and (C) The original visualization is converted to an accessible one by a singe-line union operation which can be executed at scale.}
    \Description[]{(A) A original line visualization that is inaccessible for people with intellectual and developmental disabilities;
	(B) A dot-based plot that is accessible;
	and (C) The original visualization is converted to an accessible one by a singe-line union operation which can be executed at scale.}
	\label{fig:casetransfer}
\end{figure*}

\subsection{Style Transfer to Accessible Visualizations}
\label{sec:example:styletransfer}
Visualization accessibility has been gaining increasing attention during the past years.
Many existing visualizations are inaccessible, and researchers recently started
calling for scalable and affordable approaches that automatically convert charts into accessible designs~\cite{kim2021accessible}.
For example,
Wu~\ea~\cite{wu2021understanding} suggested that the commonly used line charts (\autoref{fig:casetransfer} (A)) are not accessible for people with intellectual and developmental disabilities,
while a dot-based plot is a better alternative (\autoref{fig:casetransfer} (B)).
One possible solution to the above problem is style transfer,
\ie to blend the data of the original visualization (visualization A) with the visual encodings of another (visualization B).

However,
existing implementations of visualization style transfer either focus on a limited set of visualization types~\cite{qian2020retrieve, wang2019visualization} or require manual editing~\cite{harper2017converting} which is difficult and time-consuming for two reasons.
First,
visualization authors need to read and comprehend the specifications for the visualization B,
which contains advanced data transformation such as \verb+window+.
Second,
they need to merge two specifications and carefully revise the values of data-related entries (\eg \verb+field+ and \verb+groupby+) that are bound.

Our implementation provides an automated approach through the union operation (\autoref{fig:casetransfer} (C)).
This operation reduces the efforts for manual editing and can apply to inaccessible visualizations at scale.
Furthermore,
it can be extended to other forms of accessible charts (\eg colorblind-friendly palettes).

We benchmark the expressiveness of our approach by constructing a test case suite of 374 Vega-Lite specifications on the official example gallery~\cite{vegaLiteExample},
to which we refer as target style visualizations.
Those visualizations vary from simple charts to layered plots and multiple-view displays.
For each test case, 
we construct a corresponding data visualization that encodes the same number and type of data fields with the style visualization.
A test case is successful if the visual encodings of the style visualization can be applied to the data visualization.
The result shows that 264 (77.4\%) test cases are successful, which is very encouraging given the wide diversity of those official examples.

We analyze the failing cases to motivate future research.
First, 53 cases are due to specifications expressed in string formats,
\eg \verb+"calculate": "(datum.year % 10)"+. 
When updating the dataset, 
our method can not parse the string format and replace the data column (\ie \verb+year+).
The second type of errors (18 cases) is related to data values such as \verb+"extent": [2500, 6500]"+,
which might become invalid extent for the new dataset.
To address those problems,
future work might apply table transformation methods~\cite{feng2017component} to parse those data-related entities and automatically update new values.
Finally,
our method cannot handle map visualizations (5 cases) and might fail at parsing time strings (1 case).

\begin{figure}[!ht]
	\centering
    \includegraphics[width=0.48\textwidth]{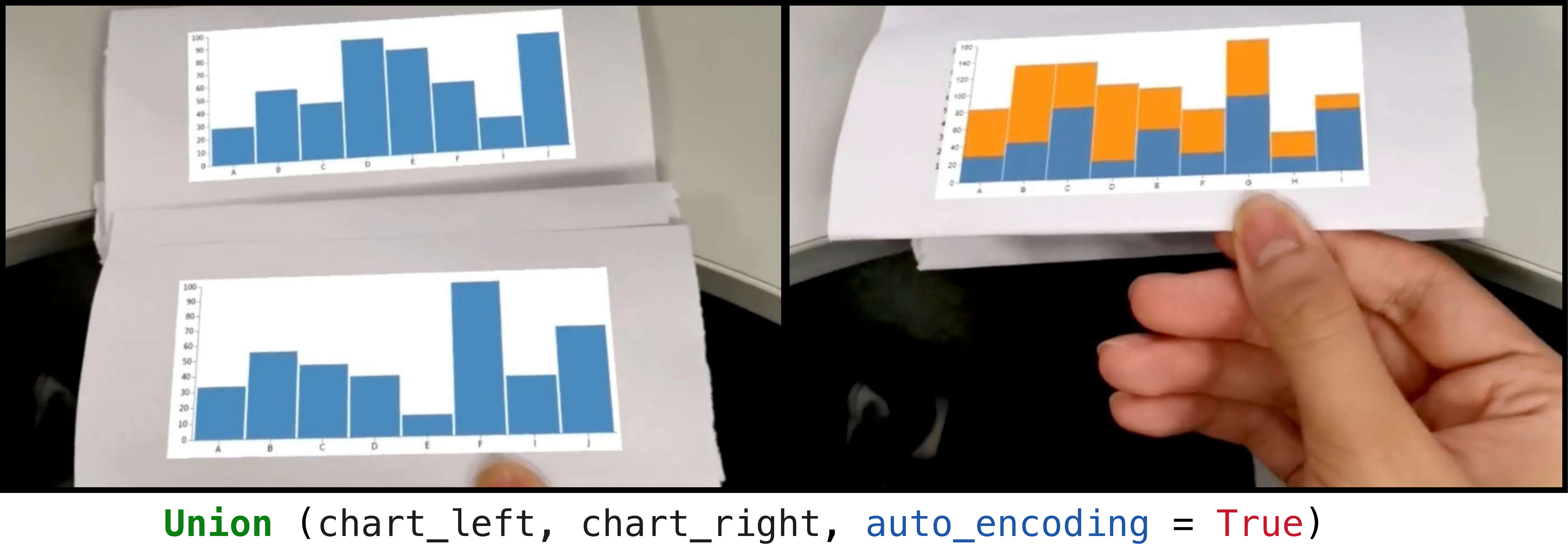}
 	\caption{With the auto-encoding mechanism, the union operation merges two bar charts into a stacked bar visualization to represent more comprehensive view of data. This example is implemented under Mobile AR environment where the union operation is performed by physically overlaying two charts.}
	  \Description[]{With the auto-encoding mechanism, the union operation merges two bar charts into a stacked bar visualization to represent more comprehensive view of data. This example is implemented under Mobile AR environment where the union operation is performed by physically overlaying two charts.}
	\label{fig:case:merge}
\end{figure}

% \begin{wrapfigure}{R}{0.5\textwidth}
%   \begin{center}
%     \includegraphics[width=0.48\textwidth]{picture/case_merge.pdf}
%   \end{center}
%  	\caption{With the auto-encoding mechanism, the union operation merges two bar charts into a stacked bar visualization to represent more comprehensive view of data. This example is implemented under Mobile AR environment where the union operation is performed by physically overlaying two charts.}
% 	  \Description[]{With the auto-encoding mechanism, the union operation merges two bar charts into a stacked bar visualization to represent more comprehensive view of data. This example is implemented under Mobile AR environment where the union operation is performed by physically overlaying two charts.}
% 	\label{fig:case:merge}
% \end{wrapfigure}

\subsection{Interactive Visualization Composition}
\label{sec:example:composition}
Visualization composition concerns combining two existing visualizations in the same visual space to facilitate visual comparison~\cite{lyi2020comparative} and more comprehensive data analysis~\cite{yang2014understand}.
Existing solutions to visualization composition such as Data Illustrator~\cite{liu2018data} are mainly implemented on graphical user interfaces, require manual manipulation or programming.
Automated methods can be helpful in reducing human efforts, increasing the scalability, and adapting to post-WIMP interfaces like AR.

% Despise some research focusing on exploring and understanding the design space of composited visualizations from a theoretical perspective (\eg~\cite{lyi2020comparative,javed2012exploring}),
% there still lacks practical implementation that assists users in compositing charts at scale.

\begin{figure}[!h]
	\centering
	\includegraphics[width=.48\textwidth]{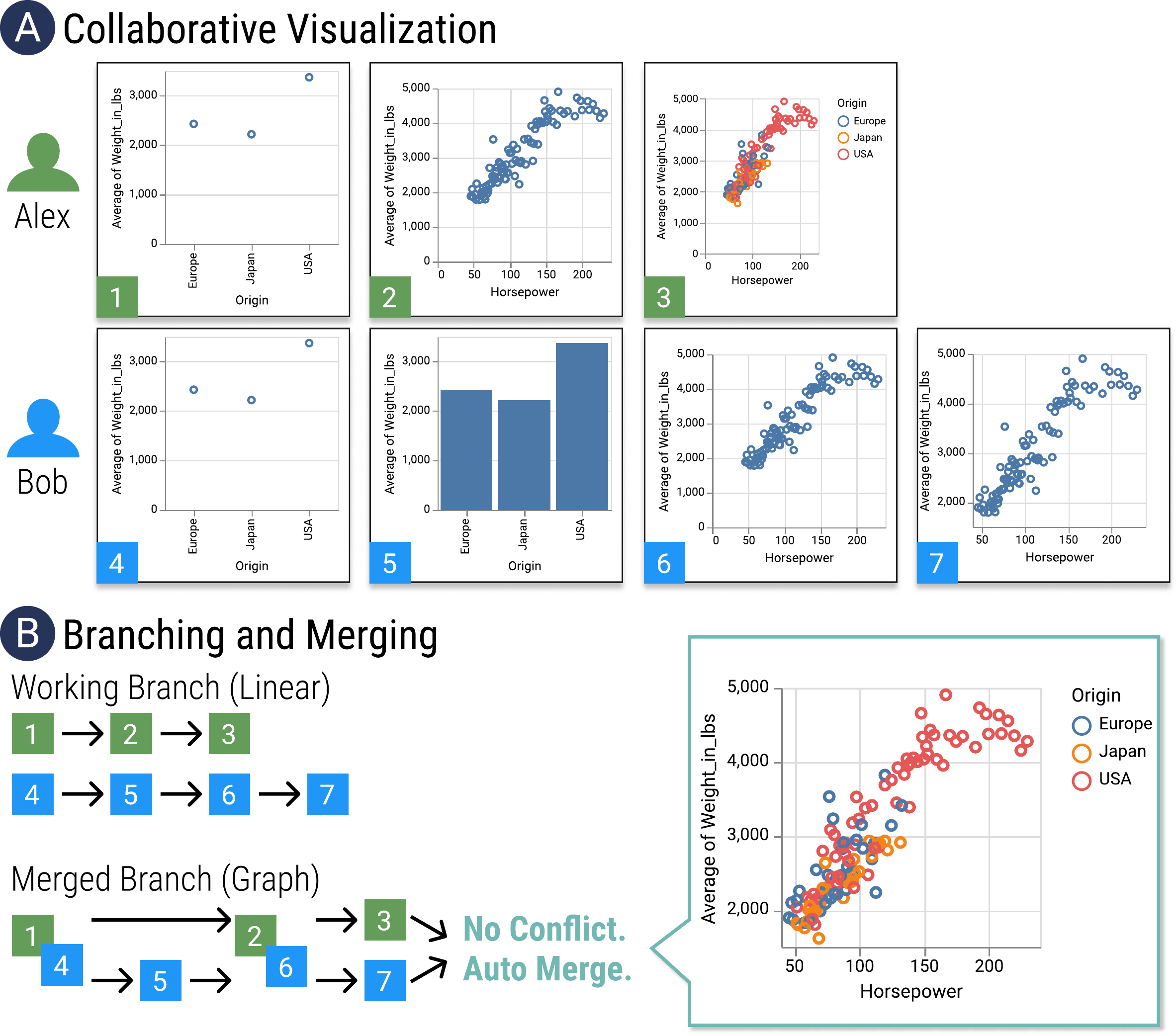}
	\caption{Handling version controls in collaborative visualization: (A) Two data workers create a series of visualization versions in collaborative visualization;
	(B) \lib{} helps merge collaborators' working branches into a graph structure. Besides, the final versions can be merged since no conflicted is detected.}
    \Description[]{(A) Two data workers create a series of visualization versions in collaborative visualization;
	(B) \lib{} helps merge collaborators' working branches into a graph structure. Besides, the final versions can be merged since no conflicted is detected.}
	\label{fig:case:git}
\end{figure}

With \lib{},
visualization composition can be implemented based on the union operator in an interactive manner.
In~\autoref{fig:case:merge},
we show an example where two standard bar charts are composited into a stacked bar visualization by overlaying them in Mobile AR environment.
This feature introduces new possibilities of visualization interactions~\cite{dimara2019interaction} by treating the whole visualization as the primary object,
which are potentially beneficial (\eg increasing user engagement) and warrant deeper studies.

We benchmark the expressiveness by replicating the composited layout in visual comparison\footnote{https://sehilyi.github.io/comparative-layout-explorer/}.
\lib{} can replicate 13 (54\%) out of the 24 cases.
Failing cases include mirror layouts (6), composite marks (3), and item-wise adjacent juxtaposition (2).
\lib{} only supports adding a new encoding channel in visualization composition,
while those cases require additional operations (\eg mirror layouts require reversing the axis of one visualization).
To solve those problems,
a promising research topic is to design and develop grammars for visualization composition (\eg~\cite{chen2021nebula}).

\begin{figure*}[!t]
	\centering
	\includegraphics[width=1\linewidth]{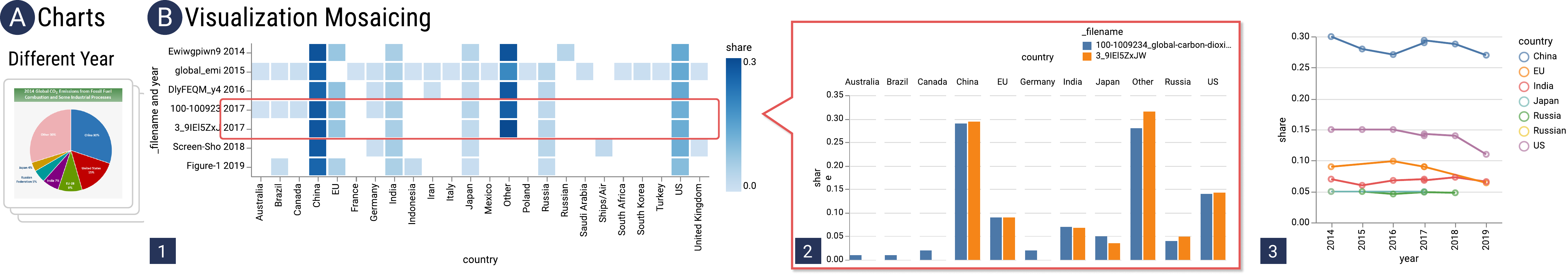}
	\caption{The union operation converts (A) multiple pie charts encoding countries' share of carbon emission in different years to (B) mosaicked visualizations including (1) a heat-map summarizing the whole data, (2) a stacked bar visualization comparing two charts with conflicting data (both in 2017), and (3) a time series showing the trend of commonly referred countries.}
    \Description[]{The union operation converts (A) multiple pie charts encoding countries' share of carbon emission in different years to (B) mosaicked visualizations including (1) a heat-map summarizing the whole data, (2) a stacked bar visualization comparing two charts with conflicting data (both in 2017), and (3) a time series showing the trend of commonly referred countries.}
	\label{fig:case:mosaicing}
\end{figure*}

\subsection{Version Control in Collaborative Visualization}

One important aspect of collaborative visualization involves version control that records changes to the visualization and resolves conflicts among collaborators~\cite{schwab2020visconnect}.
Although version control is a well-established technique in software engineering,
its adoption to data visualizations remains low. 
Specially,
existing approaches mainly use timestamps to resolve orderings that are not aware of the visualization content~\cite{badam2014polychrome} or do not support conflict detection~\cite{cutler2020trrack}.

\lib{} addresses this problem by creating a history graph recording different versions of the visualization.
For instance,
in~\autoref{fig:case:git},
the historical versions made by two collaborators are represented by branches,
where each node represents a version.
Branches can be merged to form a graph by grouping and combining identical nodes using the difference operation (\ie the difference between two charts is none).
Such graph-based representations can effective reveal the patterns such as vertex connectivity (representing commonly visited versions),
which can not be easily found in linear representations~\cite{xu2020survey}.
Besides,
\lib{} can detect merge conflicts that are common in collaboration.
As shown in \autoref{fig:case:git} (B),
the visualizations created by two collaborators (Alex and Bob) diverge after the version 2/6,
where Alex adds a new color encoding and Bob adjusts the axes' extent.
Since two changes are not conflicting (by using the difference operator),
the final visualizations can be merged to facilitate the collaboration.

\subsection{Visualization Mosaicing}
\label{sec:case:mosaicing}

As visualizations are increasingly shared on the web,
so are their impacts on the public and society.
As such,
recent research has started to investigate the information communicated through data visualizations~\cite{zhang2021mapping,comba2020data}.
However,
their approaches rely on manual analysis which does not scale well.

We introduce visualization mosaicing as an initial approach towards more automated analysis,
that is,
two or more charts of different views of a dataset can be mosaiced to represent the complete view.
We borrow this concept from similar ones such as image mosaicing~\cite{capel2004image}.

In~\autoref{fig:case:mosaicing} (A),
we show a collection of charts showing the countries' share of carbon emission in different years.
Those visualizations are collected from Google Image Search.
We leverage reverse engineering techniques~\cite{poco2017reverse} to obtain their corresponding specifications.
To get an overview of those visualizations,
we apply multiple-way union and visualize the results using a heat-map.
In~\autoref{fig:case:mosaicing} (1),
we observe how frequently each country is mentioned in original visualizations,
\ie common countries include China, India, Japan, and US which might be due to their significant emission.
Only one visualization includes data from (international) ships and air,
which might prevent the public from understanding that aviation is considerably responsible for carbon emission.
Besides,
we note that two charts encode the data in 2017.
By composing them in~\autoref{fig:case:mosaicing} (2),
we observe that their data is not consistent,
which indicates potential data errors and misinformation.
Finally,
we can convert the heat-map to a time series (\autoref{fig:case:mosaicing} (3)) to explore the trend of commonly referred countries,
which cannot be easily found in the original visualization collection.

\subsection{Computing Visualization Embedding for Clustering Analysis}
\label{sec:case:clustering}
Clustering has emerged as a new technique for analyzing a visualization collection and conducting meta-visualization analysis (\eg~\cite{zhao2020chartseer, xu2018chart}).
The state-of-the-art method~\cite{zhao2020chartseer} leverages grammar variational autoencoders (GVAE) to convert Vega-Lite specifications into embeddings,
which are subsequently projected onto a two-dimensional space for clustering analysis.
However,
the reported performance of GVAE remains relatively low (\ie 44.46\%) which degrades the final clustering results.
Besides,
such methods are end-to-end and not configurable,
while users might want to specify criteria for visualization clustering (\eg cluster charts by types or colors).

To solve those problems,
we propose visualization clustering based on \lib{}.
Specifically,
we first execute the difference operation to derive the results (\autoref{fig:example_operation} (B)).
We then apply a manual-configurable weighting function to different properties,
whereby obtaining a score measuring the distance between two visualizations.
The above steps are repeated for every pair of visualizations to get a distance matrix,
which is subsequently fed into a multidimensional scaling model~\cite{cox2008multidimensional} to compute the visualization embeddings.
Critically,
the weighting function is configurable, 
which could be designed for particular purposes based on domain knowledge or integrate cost functions from previous research such as GraphScape~\cite{kim2017graphscape}.

\begin{figure}[!t]
	\centering
    \includegraphics[width=0.48\textwidth]{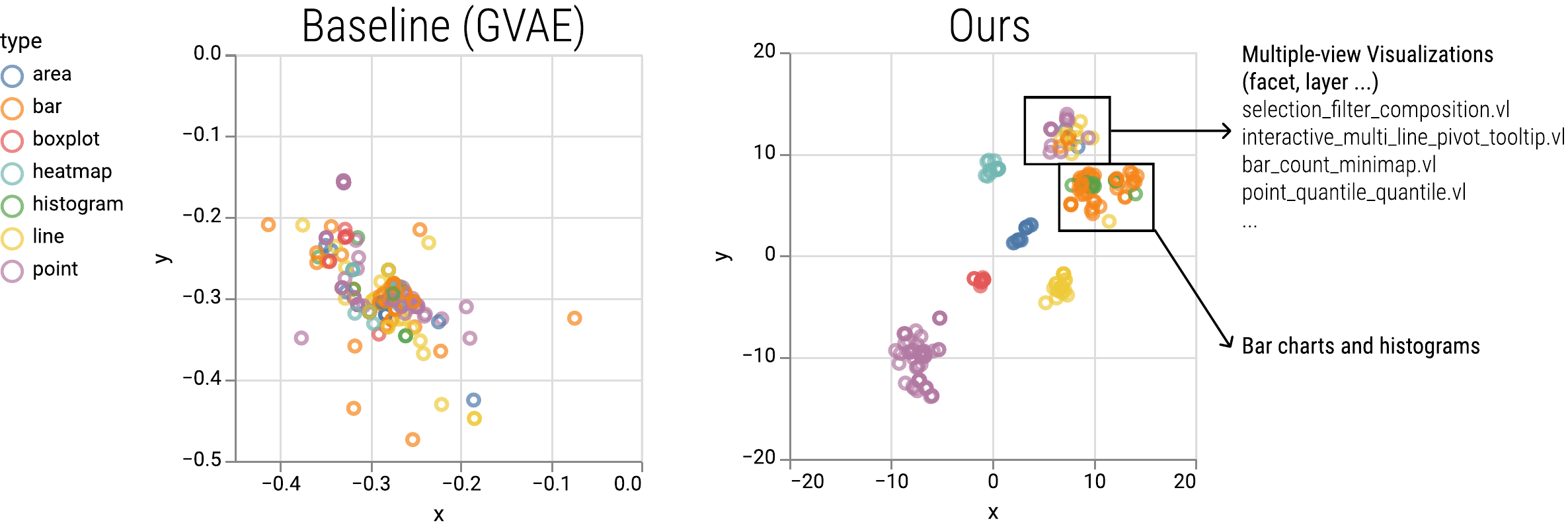}
	\caption{Visualizations of the learned embeddings by the baseline and our approach.}
    \Description[]{Visualizations of the learned embeddings by the baseline and our approach.}
	\label{fig:case:clustering}
\end{figure}

% \begin{wrapfigure}{R}{0.5\textwidth}
% 	\centering
% 	\includegraphics[width=.48\textwidth]{picture/case_clustering.pdf}
% 	\caption{Visualizations of the learned embeddings by the baseline and our approach.}
%     \Description[]{Visualizations of the learned embeddings by the baseline and our approach.}
% 	\label{fig:case:clustering}
% \end{wrapfigure}

We conduct an experiment of clustering visualization by their types.
The dataset includes Vega-Lite specifications in the official example gallery~\cite{vegaLiteExample},
where the ground truths of visualization types are available in the file name.
We filter out visualization types whose frequency is less than ten and obtain 210 visualizations in total.
According to our domain knowledge,
we assign higher weights to \verb+mark_type+ (the type of visual marks) and \verb+field_type+ (the data type of encoding channels) that are closely related to visualization types.
We compare our method with the GVAE algorithm in ChartSeer~\cite{zhao2020chartseer} and visualize the embeddings using t-SNE~\cite{van2008visualizing}.
As shown in~\autoref{fig:case:clustering},
our method can separate visualizations by their classes,
while the baseline approach achieves poorer performance.
Visualizations embeddings of bar charts and histograms are projected nearby, which is reasonable.
Besides,
we note that multiple-view visualizations are clustered together irrespective of the visualization type.
In summary,
our method demonstrates a feasible and effective solution for clustering visualizations by customizable criteria.

\subsection{Exploring the Genealogy of Visualizations}
The large collection of online visualizations has served as a source for visualization designers and researchers to explore and investigate design demographics. 
Existing work is mainly based on simple statistics to explore questions such as ``which visual attribute---position, length or color---is most commonly used to encode data?''~\cite{hoque2019searching, battle2018beagle}.
A missing perspective is the relationships among those designs,
which is important for understanding the design evolution.

\begin{figure}
	\centering
	\includegraphics[width=.5\textwidth]{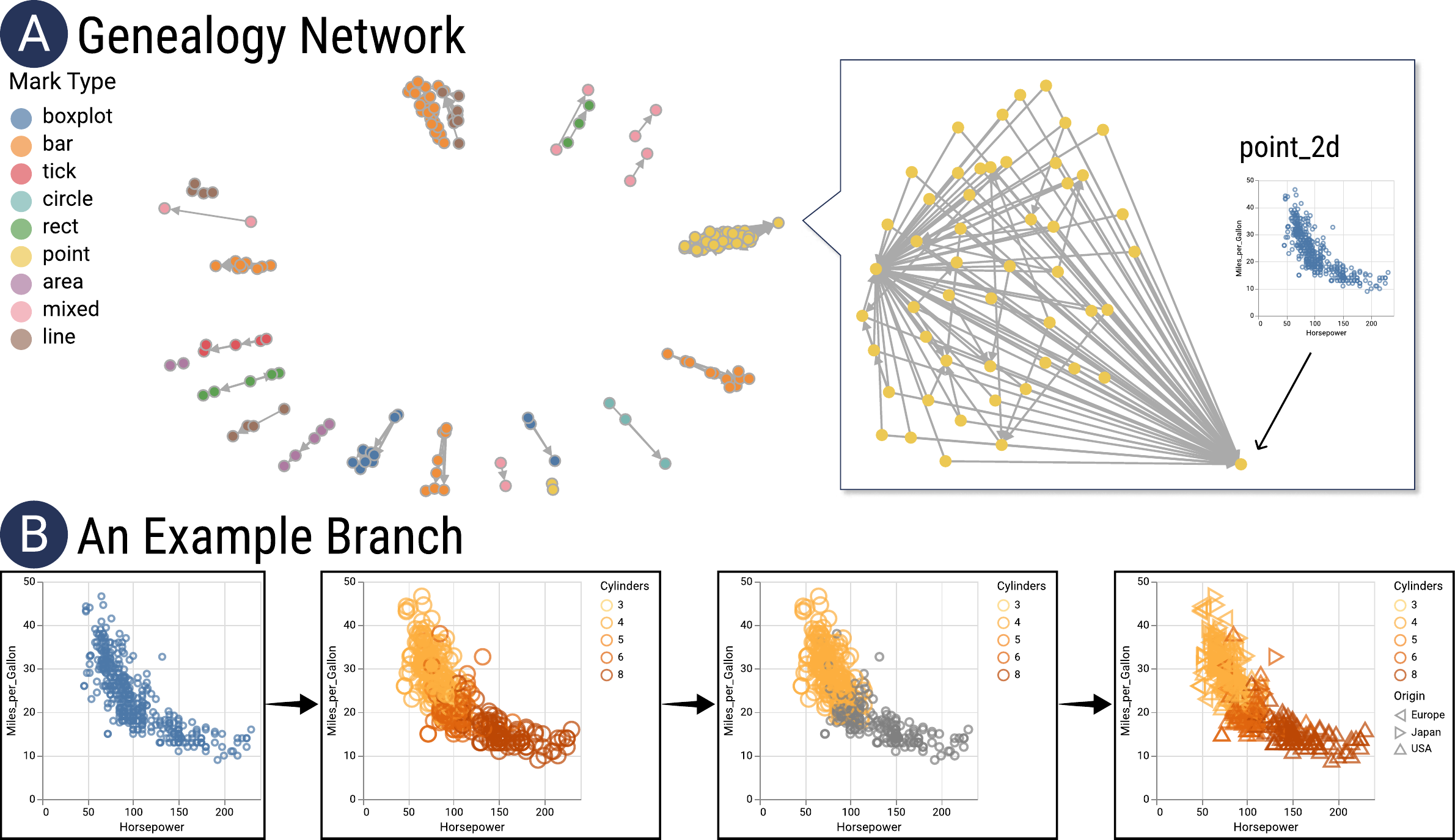}
	\caption{Analysis of the ``genealogy'' of visualizations: (A) The genealogy tree of 340 Vega-Lite examples; (B) A example branch showing how a scatterplot design evolves to encode more information or enable new interactions.}
    \Description[]{(A) The genealogy tree of 340 Vega-Lite examples; (B) A example branch showing how a scatterplot design evolves to encode more information or enable new interactions.}
	\label{fig:case:genealogy}
\end{figure}

We construct the genealogy network of the Vega-Lite examples~\cite{vegaLiteExample}.
Specially,
we define that a visualization is an ancestor if it is a subset of another visualization.
We compute such relationships by the intersection operation and ignore specification entities that are irrespective of visual designs such as \verb+description+. 

\autoref{fig:case:genealogy} (A) provides an overview of the derived genealogy network with dozens of sub-networks.
We note a large sub-network consisting of yellow-colored nodes,
which is rooted from a simple 2D scatterplot.
This shows that the scatterplot has evolved into a wide range of more complex designs.
\autoref{fig:case:genealogy} (B) shows an example branch of design evolution.
The genealogy network allows us to quickly find a simpler and more advanced version of visualization designs.
Furthermore,
the network can be used to organize the tutorials for specification learners.
For example,
the genealogy branch can be converted to a step-by-step tutorial with increasing complexity.
When learners have difficulties in understanding complex visualization specifications,
they could traverse to the ancestor as prerequisites.
Finally,
siblings can be recommended as ``related topics'' to guide exploration.

% \yong{Reach here.}

\section{Discussion and Future Work}
The past years have witnessed a growing body of research on processing and analyzing visualizations after they have been generated and digitized,
with substantial advances being made on developing new techniques (\eg visualization sequencing~\cite{kim2017graphscape}) and applications (\eg style transfer~\cite{qian2020retrieve}).
Our work is motivated by a gap concerning the isolation among research ideas
and we aim to contribute to an improved understanding and clearer organization of existing literature.
We therefore propose defining mathematical operations as a formalism for processing multiple visualizations.
In this section,
we discuss the capabilities and limitations of our formalism to inform future research.

% Our work contributes to a new way of thinking about mathematical operations as a formalism for processing multiple visualizations.
% This formalism allows us to abstract a variety of visualization-related tasks into low-level data operations (\autoref{fig:operator_overview}),
% which contributes to an improved understanding and clearer organization of existing research work.

\subsection{Descriptive and Generative Power}
We now discuss the utility of our formalism regarding its descriptive power (\ie to describe existing solutions) and generative power (\ie~to help create new ideas and solutions).

The descriptive power \begin{revised}is rooted\end{revised} in our classifications for operations in terms of \textit{what} (operation target) and \textit{how} (operation type).
Our formalism is therefore compact and expressive - basic operations can describe complex tasks, \eg style transferring is described by a union operation on data and style.
As illustrated in~\autoref{sec:ope:basic},
the proposed formalism enables us to clearly express, classify, and organize relevant work at the conceptual level.
From a practical perspective,
we demonstrate its ability and potentials to facilitate code reuse despite the disparities between the applications.
For instance,
the union operation in our implementation can reproduce both style transfer (\autoref{sec:example:styletransfer}) and visualization composition (\autoref{sec:example:composition}),
which are previously separated.
However,
we acknowledge that our formalism cannot represent operations directly on visualization images such as transmogrification~\cite{brosz2013transmogrification}.
Future work could extend our formalism to support such tasks.

Much of the generative power originates from the starting formalism based on binary operators.
The formalism is thus compositional and modular,
as binary operations are the keystone of advanced computations concerning more than two visualizations and even a collection of visualizations.
As discussed in~\autoref{sec:operation:valid:advanced},
we identify six advanced operations using binary operations as a basis.
For example,
both filtering and clustering build upon measuring the distance between two visualizations,
which can be implemented by an intersection operator combined with an additional weighting function (\autoref{sec:case:clustering}).
Although our current advanced operators might not be comprehensive,
our formalism can point researchers toward other scenarios where new operators can be generated.
Furthermore,
we show that our formalism can generate actionable solutions to emerging problems such as creating accessible visualizations (\autoref{sec:example:styletransfer}).
We hope that our formalism could help abstract other problems in terms of which and how operators can be applied.

% Our work highlights the novel perspective of thinking and formalizing visualizations as computable data.
% This perspective allows us to abstract a variety of visualization-related tasks into fundamental data operation (\autoref{fig:operator_overview}),
% which contributes to a deeper understanding and clearer organization of existing research questions.
% We demonstrate that \lib{} facilitates a range of tasks for processing and analyzing existing visualizations in a more automatic and efficient manner.
% In addition to the existing use cases of \lib{},
% there are existing potentials for new research problems and applications that warrant future work.

% However, I am not entirely convinced if the proposed approach in ComputableViz is solid in terms of usability and generalizability. One of the main concern I have is that it looks many parts of the system is relying on heuristics (e.g., inferring key columns) and manual configurations (e.g., determining which properties of the spec should be chosen from two visualizations). Which means, if the visualization becomes more complex, the system might fail to work properly or require modifications that are time-consuming. Visualizations that are used in the case studies are reasonably basic (e.g., bar charts, line charts, scatterplots), and benchmark results show that many cases are actually failing (i.e., Section 5.1-2).

\subsection{Scalability and Usability}
Our proof-of-concept implementation provides reliable starting points for performing mathematical operations on visualization specifications to enable new applications.
There remain several challenges and promising research opportunities for improving the scalability and usability of~\lib{}.

\textbf{Improving the heuristics.}
The current implementation relies on several heuristics,
\eg to infer key columns and detect data conflicts.
Those heuristics are simple, manageable, and efficient in terms of running time.
However,
heuristics can be error-prone,
especially when applying to real-world visualizations with considerable data noises,
\eg~in~\autoref{sec:case:mosaicing}.
To address this problem,
future research might leverage machine learning methods (\eg~\cite{nargesian2018table}) to improve the performance.
However,
this perspective presents significant research challenges such as gathering high-quality training data, adopting and evolving models, and improving the model explainability and trustworthiness~\cite{saket2018beyond}. 

\textbf{Refining the programming abstraction.}
Our implementation inherits the abstraction from the conceptual formalism (\autoref{fig:operator_overview}),
\eg~it abstracts both style transfer and visualization composition into a single union function through parameterization.
Although this abstraction generalizes behaviors and promotes code reuse,
it could have a cost related to usability and scalability.
For example,
end-users need to manually configure those parameters,
\eg~determining which properties of the specifications should be chosen.
Such manual configuration can be diminished by implementing specific functions for style transfer and visualization composition, respectively,
but at the cost of smaller scope of customization and flexibility.
Given the practical nature of software engineering,
future research should survey end-users in the wild to refine the programming abstraction.

\textbf{Representing visualization data more expressively.}
Our benchmark in~\autoref{sec:chartPA} shows that our relational representations of visualization produce a few failing cases,
especially for non-basic visualizations such as composited and layered charts.
For example,
it cannot handle expressions in the string format for custom calculations such as \verb+"calculate": "datum.Time/datum.TotalTime * 100"+,
which requires new approaches to parse strings into relational formats.
Besides,
the union operation can not automatically produces mirror layouts and composite marks but needs tedious manual adjustment.
A potential approach is to develop more expressive language for visualization composition.
Recent work on coordinated grammars for graphics~\cite{chen2021nebula} seems quite promising and is worth further exploration.

\subsection{Generalizability and Interoperability}
This work treats Vega-Lite specifications as the primary data representation of visualizations.
A clear next step is to extend \lib{} to other types of representations such as other visualization languages and images.
In the following text,
we discuss the current limitations and improvement roadmap regarding generalizability and interoperability.

\textbf{Generalizing to other visualization language.}
An important step of \lib{} is to convert Vega-lite specifications to relational representations for computation.
This conversion is generalizable to other declarative language such as Vega~\cite{satyanarayan2015reactive} and ECharts~\cite{li2018echarts}.
However,
it is more challenging to apply \lib{} to imperative language such as D3~\cite{bostock2011d3} which has higher flexibility and expressive power.
Thus,
it would be beneficial to develop new methods for translating among different visualization language to promote interoperation.
%\eg with deep learning~\cite{roziere2020unsupervised}.

\textbf{Continuing research on converting visualization images to specifications.}
Visualizations are often created and shared in image format.
To apply \lib{} on visualization images,
it is necessary to extract the specifications from images through computer vision techniques (\eg~\cite{poco2017reverse}).
However,
those techniques are not sufficiently robust~\cite{davila2020chart}.
In our use case (\autoref{sec:case:mosaicing}),
we manually fixed the errors in extracted specifications,
which could become less affordable with the increasing number of visualizations.
Therefore,
it is necessary to continue research and improve performance.

% \subsection{Visualizations as Data about Data}
% Visualizations are traditionally described as graphical representations of data.
% We extend this definition by considering visualizations themselves as data,
% \ie data that provides information (visual encodings) about other data.
% This perspective allows us to abstract a variety of visualization-related tasks into fundamental data operation (\autoref{fig:operator_overview}),
% which contributes to a deeper understanding and clearer organization of existing research questions.
% Moving forward,
% the ``visualization as data'' perspective opens up new research opportunities.

% \textbf{Revisiting classic data problems.}

\subsection{Towards Visualization Analysis}
Our work contributes initial efforts toward automated analysis of existing visualizations.
With the rapid proliferation of data visualizations online,
visualizations are becoming an increasingly important medium for communicating information,
comparable with other mediums such as text and images.
This perspective underscores the promising opportunities of research on visualization analysis akin to image analysis and text analysis.

\textbf{Exploring the power of visualizations on the web.}
We present an application in \autoref{sec:case:mosaicing} where \lib{} allows summarizing visualizations and identifying potential data errors.
There exist huge potentials for other applications by making use of the huge number of visualizations online.
For example,
data visualizations often appear in web articles and social media and have strong persuasive power and impact~\cite{pandey2014persuasive}.
How can we analyze those visualizations to understand their content, semantics, and indented messages especially in high-impact domains such as public health and politics?
How can we automatically detect visualizations with misinformation or harmful content?
Besides,
visualizations are often byproducts in data-enriched documents such as financial statements and Wikipedia,
containing extensive information and human knowledge.
It would be helpful to mine visualizations to improve our understandings of those documents.

\textbf{Proposing new techniques for analyzing visualizations.}
We propose mathematical operations on data visualizations,
which provide a foundation for analysis and have much room for future improvement.
First,
\lib{} implements basic operators, 
based on which users could create ad-hoc solutions to advanced operators (\eg clustering).
An important extension of \lib{} is to build a collection of algorithms for visualization analysis,
that is similar to image analysis libraries such as scikit-image~\cite{scikit}.
Second,
existing algorithms for converting visualization images to specifications are imperfect,
producing errors and uncertainty.
Thus,
future techniques for visualization analysis need to be uncertainty-aware.
A potential solution is to involve human in the loop for conducting analysis,
\eg by visual analytics approach.
Finally,
visualizations are characterized by multi-modality including images, text, and data~\cite{wu2021ai4vis},
posing unique research challenges for visualization analysis.
This distinction calls for the needs of new techniques that are more tailored to visualizations.

\textbf{Thinking visualizations as data about data.}
Visualizations are traditionally described as graphical representations of data,
while recent research has started to think visualizations themselves as data,
\ie data that provides information (visual encodings) about other data~\cite{wu2021ai4vis}.
% We revisit this definition by considering visualizations themselves as data,
% \ie data that provides information (visual encodings) about other data.
Without knowing the underlying encoded data of visualizations, it is very difficult to leverage their content, especially in combination with others~\cite{wu2021multivision}.
We apply heuristic-based approaches to detect value conflicts and merge data tables,
which has limited applicability in real-world scenarios with noisy datasets.
For example,
it is necessary to link the data entities in visualizations that might have different wordings.
In the future,
we plan to integrate the techniques for analyzing data tables in the database and web research such as entities linking~\cite{limaye2010annotating} and table union search~\cite{nargesian2018table}.
We hope that such interdisciplinary perspective will inspire and engage the research community to develop innovative techniques and applications for analysing and mining visualizations.

\textbf{Building a closed-loop ecosystem of online visualizations.}
Our work is primarily concerned with processing and analyzing multiple visualizations after they have been generated and digitized.
Such process and analysis should not be an endpoint,
but in turn can empower us to create new and better visualizations.
For instance,
we show in~\autoref{sec:example:styletransfer} that reusing visualizations enables us to create accessible visualizations in a scalable manner.
Moving forward,
we plan to
explore how our formalism could foster the creation of more advanced visualizations such as multiple views, nested and composited visualizations.
To that end,
it would be beneficial to visit other formalisms or design space in related areas such as multiple coordinated views~\cite{weaver2004building,knudsen2016view} and visualization construction~\cite{mendez2016ivolver,mei2018design}.
This perspective poses new challenges on the extension of our formalism to support operations on additional factors such as layout and inter-view relations.

% \textbf{Interweaving with database and web research.}
% An important perspective on the relationships between visualizations is the encoded dataset.

% To understand the ,
% an important task is to 

% An important task of 

% We apply  determine the relationships between two data tables encoded in visualizations.

\section{Conclusion}
In this paper we propose mathematical operations as a formalism for operating multiple visualizations.
Formalising visualizations as computable data prompts us to abstract a wide range of visualization-related tasks into basic data operations
and empowers us to develop a proof-of-concept library for visualization operations.
Through several use cases,
we demonstrate that visualization operations provide feasible solutions to emerging questions regarding the process and analysis of visualizations.

We offer our formalism as one potential method for modeling visualizations as computable data,
which is still an initial step towards this direction.
% should be viewed as a very initial step.
Given the rapid proliferation of visualizations online,
we expect that visualization processing and analysis will attract extensive research interests from not only the visualization and HCI community but also other related fields such as database, computer vision, and the web.
We hope that our work will inspire researchers to think broadly about how to advance the management and analysis of the massive data visualizations on the web.

% Given the rapid proliferation of visualizations online and the interdisciplinary nature of visualization analysis,

% We highlight that such research involves interdisciplinary research

% we introduce mathematical operations as a new abstraction for processing and analysing existing data visualizations.
% We develop a proof-of-concept library and demonstrate its usefulness and expressiveness through several use cases.
% We also validate our design through a systematic review of existing work and discuss future research opportunities.
% By opening up the new perspective that formalizes visualizations as computable data,
% our work has the potential to inspire research regarding the management and analysis of data visualizations.

\begin{acks}
We thank reviewers for their thoughtful suggestions.
% The research is partially supported by Hong Kong RGC GRF grant 16213317. 
\end{acks}

\bibliographystyle{ACM-Reference-Format}
\bibliography{main}

\appendix

\end{document}